\documentclass[notitlepage, showpacs, nobalancelastpage, twocolumn, aps,prb,superscriptaddress]{revtex4-2}
\usepackage[english]{babel}
\RequirePackage[T1]{fontenc}
\RequirePackage{times}
\usepackage{pifont}
\usepackage{siunitx}
\usepackage[italicdiff]{physics}
\usepackage{amsfonts}
\usepackage{subfigure}
\usepackage{amsmath}
\usepackage{amssymb}
\usepackage{graphicx, epstopdf}
\usepackage{appendix}
\usepackage{xspace}
\usepackage{lipsum}
\usepackage{array}
\usepackage{hyperref}
\usepackage[dvipsnames]{xcolor}
\usepackage{my_symbols}
\usepackage{physics}
\usepackage{CJK}
\usepackage{bm}
\usepackage{tikz}
\usepackage{booktabs}
\usepackage{multirow}
\usepackage{pifont}
\usepackage{setspace}

 \usepackage{esvect}

\begin{document}
\begin{CJK*}{UTF8}{gbsn}
	\title{Antiferromagnetic skyrmion as a magnonic lens}

  \author{Hongbin Wu (武宏斌)}
	\affiliation{Department of Physics, School of Science, Tianjin University, 92 Weijin Road, Tianjin 300072, China}
	\affiliation{Center for Joint Quantum Studies, School of Science, Tianjin University, 92 Weijin Road, Tianjin 300072, China}
	\affiliation{Tianjin Key Laboratory of Low Dimensional Materials Physics and Preparing Technology, Tianjin University, Tianjin 300354, China}
	\author{Zi-Wu Wang (王子武)}
	\email[Corresponding author:~]{ziwuwang@tju.edu.cn} 
	\affiliation{Department of Physics, School of Science, Tianjin University, 92 Weijin Road, Tianjin 300072, China}
	\affiliation{Tianjin Key Laboratory of Low Dimensional Materials Physics and Preparing Technology, Tianjin University, Tianjin 300354, China}
	\author{Jin Lan (兰金)}
	\email[Corresponding author:~]{lanjin@tju.edu.cn} 
	\affiliation{Department of Physics, School of Science, Tianjin University, 92 Weijin Road, Tianjin 300072, China}
	\affiliation{Center for Joint Quantum Studies, School of Science, Tianjin University, 92 Weijin Road, Tianjin 300072, China}
	\affiliation{Tianjin Key Laboratory of Low Dimensional Materials Physics and Preparing Technology, Tianjin University, Tianjin 300354, China}
	
	\begin{abstract}
		A lens, a device transforming propagation directions in an organized fashion, is one of the fundamental tools for wave manipulation.
    Spin wave, the collective excitation of ordered magnetizations, stands out as a promising candidate for future energy-saving information technologies.
		Here we propose  theoretically and verify by micromagnetic simulations,  that an antiferromagnetic skyrmion naturally serves as a lens for spin wave, when the Dzyaloshinskii-Moriya strength exceeds a threshold.
		The underlying mechanism is the spin wave deflection caused by Dzyaloshinskii-Moriya interaction, a mechanism that is ordinarily overshadowed by the magnetic topology.
  	\end{abstract}
	\maketitle
\end{CJK*}

\section{Introduction}

The propagation direction, which conveys the information flux of a wave, constitutes a fundamental degree of freedom common to all wave types, including light \cite{arbabi_Advances_2023}, sound \cite{assouar_acoustic_2018}, water wave \cite{zhu_controlling_2024}, seismic wave \cite{ben-menahem_seismic_1981}, and spin waves \cite{stancil_spin_2009}. 
Consequently, achieving precise control over this directional degree of freedom represents a central objective in magnonics \cite{chumak_magnon_2015,pirro_advances_2021,flebus_2024_2024}, a field dedicated to spin-wave manipulation. 
Versatile schemes in controlling spin-wave propagation have emerged, including guiding \cite{lan_spinwave_2015,wagner_magnetic_2016, sluka_emission_2019,wang_inversedesign_2021}, reflection \cite{yu_magnetic_2016, yu_electromagnetic_2026}, refraction \cite{mulkers_tunable_2018, lan_geometric_2021} and deflection \cite{zhang_magnon_2023,liang_asymmetric_2024}.

A quintessential  device for controlling propagation direction is the magnonic lens, which interconverts spin waves between point-source and plane-wave forms. 
Such lenses are typically realized by creating magnetic inhomogeneity, either through the introduction of material boundaries~\cite{bao_offaxial_2020} or by engineering spatial gradients in saturation magnetization~\cite{davies_Gradedindex_2015, dzyapko_reconfigurable_2016, gruszecki_Spinwave_2018, whitehead_luneburg_2018, whitehead_Graded_2019, vogel_optical_2020}. 
The underlying physics is a magnetic analog of Snell's law~\cite{stigloher_snells_2016,yu_magnetic_2016, mulkers_tunable_2018, hioki_snell_2020}, which describes the relationship between incident and refracted paths at magnetic interfaces.

Beyond such deliberately engineered systems, naturally emerging magnetic textures offer a complementary pathway for spin-wave control \cite{yu_magnetic_2021,mochizuki_thermally_2014,saji_hopfiondriven_2023}, leveraging their inherent topological protection \cite{nagaosa_topological_2013,zang_topology_2018}. 
However, the manipulation of fast spin waves by these slowly evolving textures involves a complex spatiotemporal interplay that is not yet fully understood, often necessitating simplified theoretical treatments. 
A further complication arises from the Dzyaloshinskii-Moriya (DM) interaction widely existing  in  inversion symmetry broken systems \cite{yang_firstprinciples_2022}, which introduces chirality to magnetic textures \cite{gobel_skyrmions_2021} and non-reciprocity to spin waves \cite{moon_spinwave_2013,wang_chiral_2020,kuepferling_measuring_2023}. 
Despite its central role, influences of DM interaction on spin wave dynamics in inhomogeneous texture backgrounds are frequently overshadowed by the impacts of magnetic topology.

\begin{figure}[b]
	\centering
	\frame{\includegraphics[width=0.5 \textwidth]{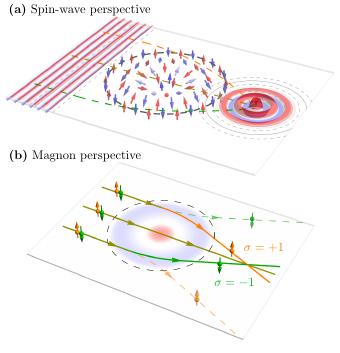}}
    \caption{
		{\bf Schematics of a skyrmion-based magnonic lens.}
		(a) Spin wave transformation through an antiferromagnetic skyrmion.
        Red and blue arrows denote two sublattice magnetizations of an antiferromagnetic skyrmion, and stripes indicate the spin wave with solid/dashed lines depict the isophase lines.   
        (b) Magnon trajectories across pseudo-magnetic field induced by the magnetic skyrmion.
        The background blue/red colors encode the pseudo-field of opposite polarities induced by the skyrmion. Orange/green lines represent the magnon trajectories with pseudo-charges of sign $\sigma=\pm 1$.
        \label{fig:lens_scheme}
        }
\end{figure}

In this work, we address this gap by establishing a correspondence between spin-wave scattering by magnetic textures and magnon deflection by a pseudo-magnetic field, thereby clearly distinguishing and quantifying the respective contributions of magnetic topology and the Dzyaloshinskii-Moriya (DM) interaction to spin-wave dynamics. Leveraging the wave-particle duality \cite{lan_skew_2021,wu_spin_2025a}, we demonstrate that an antiferromagnetic skyrmion can act as a magnonic lens by partially converging incoming spin waves, as schematically illustrated in Fig.~\ref{fig:lens_scheme}, provided the DM interaction strength exceeds a threshold value. Our analysis identifies the pseudo-magnetic field induced by the chiral DM interaction, rather than the field originating from magnetic topology, as the key mechanism responsible for spin-wave focusing. By utilizing an antiferromagnetic skyrmion as a natural magnonic lens, we establish a new paradigm for harnessing magnons through the DM interaction.

\section{Basic model}
\subsection{Antiferromagnetic skyrmion}

Consider an antiferromagnetic film extending in the $x$-$y$ plane, where the sublattice magnetizations $\mb_{1}$ and $\mb_{2}$ are depicted by red and blue arrows in Fig.~\ref{fig:lens_scheme}. Following conventions in antiferromagnets, we define the staggered magnetization $\bn  \equiv  \qty(\mb_{1} - \mb_{2})/2$ and net magnetization $\mb \equiv (\mb_{1} + \mb_{2})/2$, subject to the unity constraint $|\bn| = 1$ and the orthogonal constraint $\bn \cdot \mb = 0$. 
The dynamics of staggered magnetization $\bn$ are governed by the antiferromagnetic Landau-Lifshitz-Gilbert (LLG) equation \cite{baryakhtar_nonlinear_1980, haldane_nonlinear_1983, tveten_antiferromagnetic_2014, wu_curvilinear_2022, liu_geometric_2023a}:
\begin{align}
	\label{eqn:LLG}
	\rho \bn \times \ddot{\bn} = \bn \times \gamma \bh + \alpha \bn \times \dot{\bn},
\end{align}
where $\ddot{\bn} \equiv \partial_{t}^{2}\bn$, $\dot{\bn} \equiv \partial_{t}\bn$, $\gamma$ denotes the gyromagnetic ratio, and $\alpha$ is the Gilbert damping constant. The effective field acting on the staggered magnetization $\bn$ is
$
\bh = ({2}/{\mu_{0}M_{s}}) \qty[ A\nabla^{2}\bn + Kn_{z}\hbe_{z} - D(\hbz\times\nabla)\times\bn]
$,
where $A$ is the exchange stiffness, $K$ is the easy-axis anisotropy along $\hbe_{z}$, $D$ is the interfacial  Dzyaloshinskii-Moriya (DM) strength with $\hbz$ the symmetry-breaking direction, $\mu_{0}$ is the vacuum permeability and $M_{s}$ is the saturation magnetization. 
Dipolar interactions are neglected due to interweaving sublattice magnetizations in antiferromagnetic systems. The parameter $\rho = \mu_{0}M_{s}a^{2}/(16\gamma A)$ in \Eq{eqn:LLG} represents the antiferromagnetic inertial density, where $a$ is the lattice constant \cite{liu_geometric_2023a}.

\begin{figure*}[bt]
	\centering
	\frame{\includegraphics[width=0.99 \textwidth]{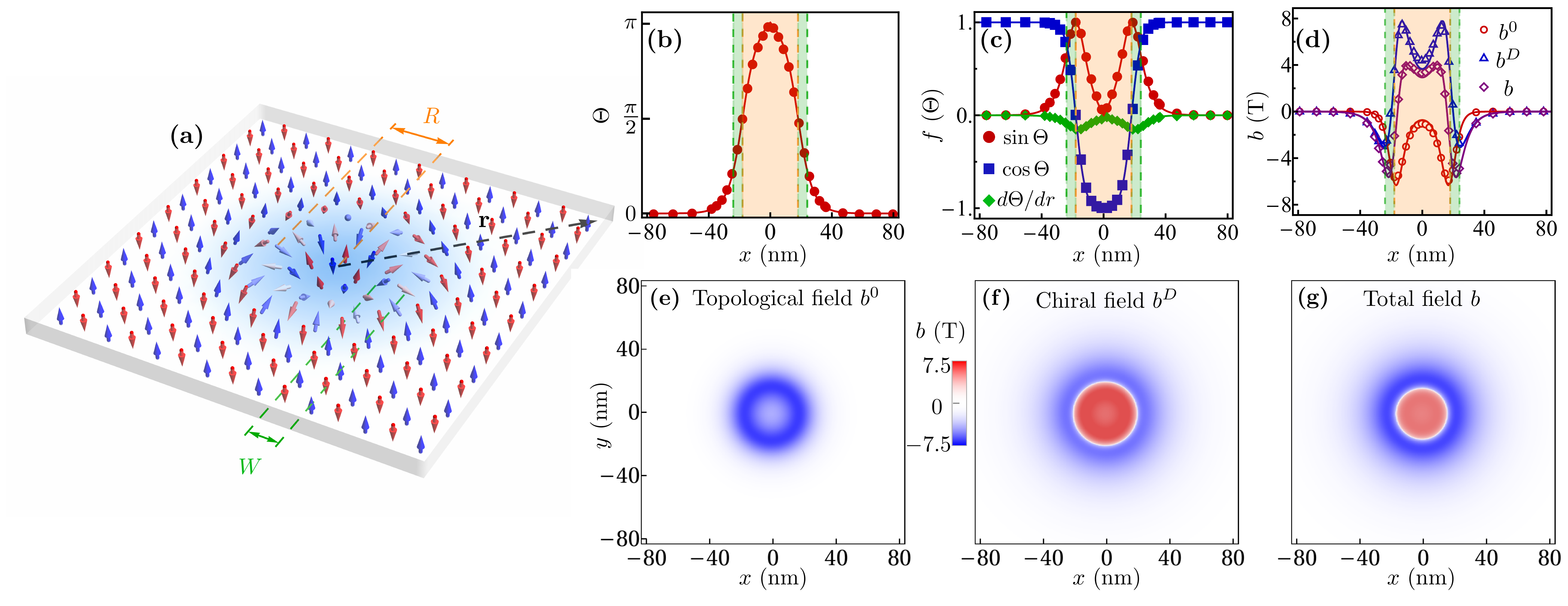}}
    \caption{
    	{\bf Spatial profiles of magnetization configuration and pseudo-magnetic field within an antiferromagnetic skyrmion}.
        (a) Magnetization distribution in an antiferromagnetic skyrmion. 
		Red and blue arrows indicate the sublattice magnetizations, and blue shading indicates the skyrmion region.  
		Black dashed lines indicate the radial direction $\br$, orange/green dashed lines represent the characteristic radius $R$ and width $W$. 
		(b-d) Cross-sectional distributions of magnetization parameter $\Theta$ and the pseudo-magnetic field.
        Lines are for theoretical modeling in \Eq{eqn:skyrmion_profile} and \Eq{eqn:b_fields}, and dots are extracted from micromagnetic simulations.
	    (e-g) Spatial distributions of the topological component $b^0$, the chiral component $b^D$ and the total pseudo-magnetic field $b$. 
		Red/blue background color encode the positive/negative values of the pseudo-magnetic field. 
        \label{fig:field_profile}
        }
\end{figure*}

Due to the chiral DM interaction, a skyrmion may form with opposite magnetization directions in its core and periphery regions \cite{nayak_magnetic_2017, gobel_skyrmions_2021, wu_Spinwave_2025a}, which resembles a $360^\circ$ domain wall in the radial direction, as illustrated in Fig.~\ref{fig:field_profile}(a).
The magnetization profile is described by 
$\bn(\br) = (\sin\Theta\cos \Phi,  \sin\Theta \sin \Phi,  \cos \Theta)$, 
where $\Theta$ and $\Phi$ denote the polar and azimuthal angles of $\bn$ relative to $\hbz$, and $\br = (r\cos\varphi,  r \sin\varphi)$ with polar coordinates $(r,\varphi)$ centered at the skyrmion core. 
Exploiting rotational symmetry, the magnetic configuration is empirically modeled as \cite{braun_fluctuations_1994,nagaosa_topological_2013,wang_theory_2018}
\begin{align}
	\label{eqn:skyrmion_profile}
	\Theta(r) &= 2 \arctan\qty(\frac{\sinh\frac{R}{W}}{\sinh \frac{r}{W}}) + \frac{1-\chi}{2}\pi, \\
	\Phi(\varphi) &= \varphi + \pi,
\end{align}
where $R$ and $W$ represent the characteristic radius and width, respectively. The polarity $\chi = \pm 1$ distinguishes two skyrmion types with inverted magnetizations, residing in domains with equilibrium magnetization $\bn_0 = \pm \hbz$.
The skyrmion radius is roughly given by \cite{wang_theory_2018}
\begin{align}
    \label{eqn:R_D}
	R= \sqrt{\frac{A}{K}\frac{D^2}{D_c^2-D^2} }, 
\end{align}
where $D_c=4\sqrt{AK}/\pi$ is the critical DM strength separating the skyrmion and spiral states \cite{rohart_Skyrmion_2013, delser_archimedean_2021}.

\subsection{ Spin wave dynamics and magnon kinetics}

When considering temporal evolution, the total staggered magnetization $\bn$ can be partitioned into a static skyrmion component $\bn_{0}$ and a dynamical spin wave component $\delta\bn$, such that $\bn = \bn_0 + \delta\bn$. For spin waves with small amplitude ($|\delta\bn| \ll 1$), the transverse condition $\delta\bn \cdot \bn_{0} = 0$ holds due to the unity constraints $|\bn| = 1$ and $|\bn_0| = 1$.  In spherical coordinates, the spin wave is described by $\delta\bn = n_\theta \hbe_\theta + n_\phi \hbe_\phi$,
where $\hbe_\theta$ and $\hbe_\phi$ are transverse unit vectors orthogonal to $\hbe_r \equiv \bn_0$. Furthermore,  the spin wave is given by the complex field:
$\psi_{\sigma} = n_\theta - i\sigma n_\phi$,
where the chirality $\sigma = \mp 1$ corresponds to left-circular ($\sigma = -1$) and right-circular ($\sigma = +1$) polarizations, respectively.

The dynamics of spin waves propagating through  an antiferromagnetic skyrmion is derived from the  LLG equation \eqref{eqn:LLG}, yielding a Klein-Gordon-like equation \cite{kim_tunable_2019,lan_skew_2021,lan_geometric_2021}:
\begin{align}
	\label{eqn:eom_sw}
	-\frac{\rho}{\gamma} \ddot{\psi}_\sigma = \dfrac{2}{\mu_{0}M_{s}} \qty[ A \left(-i\nabla + \sigma \ba \right)^2 + K ] \psi_\sigma,
\end{align}
where $\ba = \ba^{0} + \ba^{D}$ denotes the effective vector potential generated by inhomogeneous magnetization.
The topological component $\ba^0 = \bm{\Lambda} \cdot \nabla\bn_0$ arises from magnetic texture \cite{kim_tunable_2019,lan_skew_2021}, with $\bm{\Lambda} = (\hbz \times \bn_0)/(1 + n_0^z)$ representing the Dirac vector potential of a monopole; the chiral component $\ba^D = (D/2A) (\hbz \times \bn_0)$  is mediated by the DM interaction \cite{kim_tunable_2019,lan_skew_2021,lan_geometric_2021}.

Within the short-wavelength approximation, the local dispersion relation derived from \Eq{eqn:eom_sw} takes the form
\begin{align}
	\label{eqn:local_omega}
	\dfrac{\mu_{0}M_{s}\rho}{2\gamma} \omega^2(\br) = A \bk^2(\br) + K,
\end{align}
where $\hbar\bk = \hbar\qty(\bq + \sigma\ba)$ represents the canonical momentum, with $\omega$ and $\bq$ denoting the local frequency and wavevector, respectively. 
In uniform domains where $\bk \to \bq$, \Eq{eqn:local_omega} reduces approximately to the linear dispersion relation $\omega \approx c q$, where 
$c =4\sqrt{2}\gamma A /(\mu_0 M_s a)$ 
 is the effective `speed of light' characterizing the antiferromagnetic medium.

Within the framework of the eikonal approximation \cite{landau_classical_1990}, the wave equation in \Eq{eqn:eom_sw} maps directly to particle kinetics in phase space $(\br, \bq)$, where the local spin wave dispersion $\omega$ in \Eq{eqn:local_omega} functions as the effective magnon Hamiltonian. 
Through the de Broglie correspondence, the magnon mass is defined as $m = \hbar|\bk|/|\bv| =\mu_0 M_s \rho \hbar\omega/\qty(2\gamma A)$, with the group velocity $\bv \equiv \dot{\br} = \partial_{\bq} \omega =2\gamma A\bk/\qty(\mu_0 M_s \rho \omega)$. 
This transformation converts the spin wave dynamics described by \Eq{eqn:eom_sw} to \cite{lan_geometric_2021,liang_bidirectional_2023}
\begin{align}
	\label{eqn:eom_magnon}
	 m \ddot{\br} = -\sigma q_0 \bv \times \bb,
\end{align}
where $\bb = \qty(\hbar/q_0)\nabla \times \ba$ represents the emergent pseudo-magnetic field, $q_0$ is the elementary charge, and $\sigma$ corresponds to the sign of pseudo-charge. 
Above mapping from spin wave dynamics \eqref{eqn:eom_sw} to magnon kinetics \eqref{eqn:eom_magnon} effectively decomposes continuous spin wave propagation into discrete magnon trajectories \cite{wu_spin_2025a}.

\subsection{Pseudo-magnetic field}

For magnon motion across a skyrmion in a magnetic film confined to the $x$-$y$ plane, only the $z$-component of the pseudo-magnetic field is physically relevant, yielding $\bb = b~\hat{\bz}$. 
Consistent with the vector potential decomposition $\ba = \ba^0 + \ba^D$ in Eq.~\eqref{eqn:eom_sw}, the pseudo-magnetic field $b = b^0 + b^D$ comprises two distinct contributions \cite{lan_skew_2021, wu_spin_2025a}: 
(i) a topological component
\begin{align}
	\label{eqn:b0}
	b^0\qty(x, y) =\frac{\hbar}{q_0} \bn_0 \cdot \qty(\pdv{\bn_0}{x} \times \pdv{\bn_0}{y}),
\end{align} 
and (ii) a chiral component
\begin{align} 
	\label{eqn:bD}
	b^D\qty(x, y) =  \frac{\hbar}{q_0}\frac{D}{2A} \nabla \cdot \bn_0.
\end{align} 
Integrating over the entire skyrmion area, $\int b^0  dx dy = \qty(\hbar/q_0)4\pi \chi$ indicates that the topological field is governed by the skyrmion polarity $\chi$ (equivalent to the topological charge), whereas $\int b^D  dx dy = 0$ shows that $b^D$ redistributes such a topological charge.

Both fields in Eqs. \eqref{eqn:b0} and \eqref{eqn:bD}  inherit the skyrmion's rotational symmetry from Eq.~\eqref{eqn:skyrmion_profile} and can be expressed in polar coordinates as:
\begin{subequations}
	\label{eqn:b_fields}
	\begin{align}
		b^0\qty(r) &= \frac{\hbar}{q_0}\frac{\sin\Theta}{r} \dv{\Theta}{r}, \\
		b^D\qty(r) &= \frac{\hbar}{q_0}\frac{D}{2A} \left( \cos\Theta \dv{\Theta}{r} + \frac{\sin\Theta}{r} \right).
	\end{align}
\end{subequations}
According to the skyrmion profile (with $\chi=+1$ for simplicity of naration) in Eq.~\eqref{eqn:skyrmion_profile}, the polar angle $\Theta(r)$ decreases monotonically from $\Theta = \pi$ at the core ($r = 0$) to $\Theta = 0$ in the periphery ($r \to \infty$).
Consequently, the topological field $b^0$ maintains a consistent sign throughout the region, while $b^D$ undergoes sign reversal due to the change in $n_0^z \equiv \cos\Theta$ from negative in the skyrmion core ($\cos\Theta < 0$) to positive in the periphery ($\cos\Theta > 0$).

The spatial pattern of the total field $b$ is determined by the competition between the topological field $b^0$ and the chiral field $b^D$, especially near the skyrmion core.
At the skyrmion center ($r = 0$), these fields are approximated as:
\begin{subequations}
	\label{eqn:b_0d_r0}
	\begin{align}
		b^0\qty(0) &\approx -\frac{\hbar}{q_0}\qty(\frac{d\Theta}{dr})^2 \approx -\frac{\hbar}{q_0}\frac{4}{R^2} < 0, \\
		b^D\qty(0) &\approx -\frac{\hbar}{q_0}\frac{D}{A} \frac{d\Theta}{dr} \approx    \frac{\hbar}{q_0}\frac{2D}{AR} > 0,
	\end{align}
\end{subequations}
where the quadratic and linear dependencies on $d\Theta/dr$ inherit from the differential orders in Eqs.~\eqref{eqn:b0} and \eqref{eqn:bD}, respectively.
From the skyrmion profile in Eq.~\eqref{eqn:skyrmion_profile}, the polar angle gradient can be approximated as $\Delta \Theta/\Delta r = -\pi/2R$ between $\Theta = \pi$ at center ($r = 0$) and $\Theta = \pi/2$ at the waist ($r = R$), or more accurately as $d\Theta/dr = -2W/\sinh(R/W) \approx -2/R$ supposing $R\gg W$.
Consequently, the interplay between the fields in Eq.~\eqref{eqn:b_0d_r0} reduces to a comparison of two characteristic length scales: the skyrmion radius $R$ and the chiral length $\xi = 2A/D$ \cite{rohart_Skyrmion_2013,wang_magnondriven_2015}.

Notably, skyrmion radius $R$ and chiral length $\xi$ exhibit opposite dependencies on the  DM interaction strength $D$: $R$ increases with $D$ to accommodate magnetization twist along the skyrmion waist, while $\zeta$ decreases to promote faster magnetization rotation.
The competition between these two scales is marked by the identity $R=\xi$, or equivalently
\begin{align}
	\qty(\frac{D^2_c}{D_t^2}-1)\frac{D^2_c}{D_t^2} = \frac{4}{\pi^2},
\end{align}
which yields the threshold
\begin{align}
    \label{eqn:Dt}
	D_t\approx \frac{\sqrt{\pi} }{2}D_c \approx 0.88 D_c.
\end{align}
The threshold $D_t$ marks the transition between a $b^0$-dominated uniform-sign pattern and a $b^D$-dominated sign-reversal pattern for the total field $b$.

To obtain explicit magnetization and pseudo-magnetic field profiles for a magnetic skyrmion, we perform micromagnetic simulations using Mumax3 \cite{vansteenkiste_design_2014}.
The material parameters, based primarily on KMnF$_3$ \cite{barker_static_2016, liang_dynamics_2019}, are as follows:
saturation magnetization $M_s = \SI{3.76e5}{A/m}$,
exchange stiffness $A = \SI{6.59e-12}{J/m}$,
easy-axis anisotropy constant $K = \SI{1.16e5}{J/m^3}$,
DM interaction strength $D = \SI{1.1}{mJ/m^2}$,
lattice constant $a = \SI{0.418}{nm}$,
and damping constant $\alpha = \SI{1e-5}{}$.
Consistet with \Eq{eqn:b_0d_r0}, our numerical analysis focuses on skyrmions with polarity $\chi = +1$; results for $\chi = -1$ follow from symmetry.

Cross-sectional distributions of the magnetization parameter $\Theta$ and pseudo-magnetic field $b$ are shown in Fig.~\ref{fig:field_profile}(b-d).
Micromagnetic simulations agree excellently with theoretical modeling using Eqs.~\eqref{eqn:skyrmion_profile} and \eqref{eqn:b_fields}, with a skyrmion radius $R = \SI{18}{nm}$ and width $W = \SI{6}{nm}$.
Since the adopted DM strength exceeds the threshold $D > D_t$, the total field $b$ exhibits the core-shell spatial distribution characteristic of $b^D$ in Fig.~\ref{fig:field_profile}(d), rather than the uniform-sign profile of $b^0$ shown in Fig.~\ref{fig:field_profile}(c).
The $b^D$-dominated field produces a highly intricate magnon deflection pattern, contrasting with the uniform deflection under $b^0$ dominance, as detailed in the following section.

\begin{figure*}[bt]
	\centering
	\frame{\includegraphics[width=0.9 \textwidth]{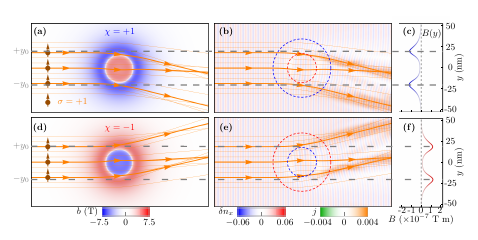}}
    \caption{
	{\bf Deflection of magnons with positive pseudo-charge $\sigma=+1$ and the corresponding scattering of right-circular spin wave by antiferromagnetic skyrmions of polarity $\chi=+1$  (a-c) and $\chi=-1$ (d-f). }
	In (a, d), red/blue color encodes the positive/negative values of pseudo-magnetic fields induced by skyrmion. 
	In (b, e), red/blue color encodes the spin wave $\delta n_x$, the orange color represents the magnitude of polarized flux $j$, and red/blue circles sketch the landscape of pseudo-magnetic field in (a, d). 
	In (c, f), red/blue lines plot  the positive/negative values of the path-integrated field  $B(y)$, respectively.
	\label{fig:one_skyrmion}
	}
\end{figure*}

\subsection{Magnon deflection}

Consider a magnon of velocity $\mathbf{v} = v_{x}\hat{\mathbf{x}}$ at height $y$ incident normally on a skyrmion. According to Eq.~\eqref{eqn:eom_magnon}, the pseudo-magnetic field deflects the magnon with polarization-dependent direction ($\sigma = \pm 1$), where the net deflection is governed by the integrated pseudo-field exposure. 
For small variations in  height, the cumulative transverse velocity $v_y$ developed during traversal is approximated as
\begin{align}
	\label{eqn:defl_vy}
	v_y \approx \int \frac{\sigma q_0 v_x b(x,y)}{m} \frac{dx}{v_x} =-\frac{q_0}{m}\sigma\chi\abs{B(y)},
\end{align}
where $B(y) \equiv \int b(x,y) dx$ represents the pseudo-magnetic field integrated along an approximately horizontal trajectory  at fixed $y$. 
The path-integrated field $B(y)$ governs the deflection magnitude, while the deflection direction is jointly determined by the polarization direction $\sigma$ and the skyrmion polarity $\chi$: the magnons are deflected clockwisely/anticlockwisely for $\sigma\chi=\pm 1$.

For the pseudo-field profile $b$ with $\chi=+1$ exhibiting core-shell pattern in Fig.~\ref{fig:field_profile}(g), the corresponding path-integrated field $B$ displays dual peaks at $|y|=y_{0}$ and a central minimum at $y=0$, as depicted in Fig.  \ref{fig:one_skyrmion}(c). 
When a right-circular spin wave of frequency $f=\SI{7}{THz}$ in plane-wave form incident on the skyrmion with polarity $\chi=+1$, equivalently a collection of magnons with pseudo-charge $\sigma=+1$ traverse on a pseudo-magnetic field with $\chi=+1$ at different heights $\{ y \}$.
According to \Eq{eqn:defl_vy}, incident magnons consequently undergo strong clockwise deflection near $y=\pm y_0$ but minimal deflection at $y=0$, and bifurcates into two bundles of rays on the right flank of the skyrmion, as demonstrated in Fig.  \ref{fig:one_skyrmion}(a). 
Correspondingly, the polarized flux $j=  (\dot{\mb}_1\times \delta \mb_1)\cdot \mb_1 - (\dot{\mb}_2\times \delta \mb_2)\cdot \mb_2 $ is extracted from micromagnetic simulations \cite{lan_geometric_2021} and plotted in Fig.   \ref{fig:one_skyrmion}(b), which exhibits nice agreement with above magnon trajectory distributions.  

Upon reversal of the skyrmion polarity to $\chi=-1$, the path-integrated field $B(y)$ changes sign, as shown in Fig.~\ref{fig:one_skyrmion}(f).
Consequently, the magnon rays switch their bending direction from clockwise to anticlockwise while preserving the dual-peak bending pattern, as depicted in Fig.~\ref{fig:one_skyrmion}(d).
The magnon trajectory distributions again exhibit good agreement with the corresponding polarized flux profile in Fig.~\ref{fig:one_skyrmion}(e).

\begin{figure*}[bt]
	\centering
	\frame{\includegraphics[width=0.98 \textwidth]{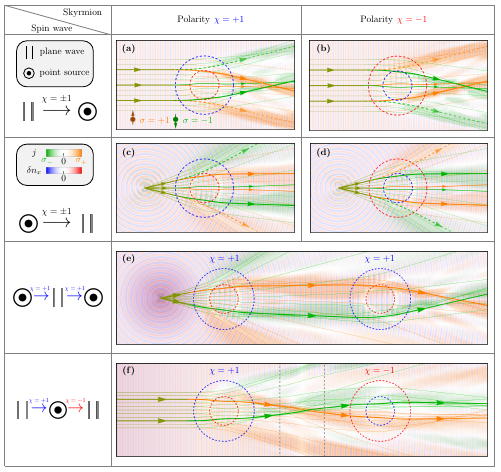}}
    \caption{
	{\bf Magnon lensing by antiferromagnetic skyrmions.}
	(a-d) Spin wave propagation modulated by a single skyrmion. 
	(e, f) Spin wave propagation modulated by a skyrmion pair. 
	In all panels, the background orange/green colors encode the spin wave fluxes of left/right-circular components extracted from the micromagnetic simulations, orange/green lines represent the magnon rays with pseudo-charge $\sigma = \pm 1$, and red/blue circles sketches the landscapes of the pseudo-magnetic fields.
	\label{fig:skyrmion_lens}
	}
\end{figure*}

The characteristic double-bending signature in Fig. \ref{fig:one_skyrmion}reveals the unique role of the chiral DM interaction in modifying magnon rays, contrasting with the ordinarily anticipated single-bending patterns produced by the magnetic topology.
Despite simple field redistribution of the topological charge, the DM interaction alters the entire magnon bending scenario in a profound fashion.

\section{Magnon lensing}

\subsection{Single magnonic lens} 

When linearly polarized spin waves (comprising of both left and right-circular components) are incident on an antiferromagnetic skyrmion  with polarity $\chi=+1$, 
they correspond to magnons with pseudo-charge of sign $\sigma = \pm 1$ traversing the pseudo-field landscape. 
The resulting deflection patterns simultaneously exhibit two superimposed versions of the double-bending shown in Fig.~\ref{fig:one_skyrmion}(a): one with downward curvature for $\sigma = +1$ and another with upward curvature for $\sigma = -1$. 
Consequently, the double-bending patterns for two circular polarizations, then gives rise to four spin wave beams, which show quantitative agreement with the spin wave fluxes in micromagnetic simulations [Fig.~\ref{fig:skyrmion_lens}(a)].
Significantly, the two inward-bending beams intersect at a focal point on the skyrmion's right flank due to their opposite deflection directions, demonstrating a magnonic focusing effect.
For a skyrmion of the opposite polarity $\chi=-1$ in \Eq{eqn:skyrmion_profile}, all magnon rays are inverted horizontally with roles of two circular polarization exchanged but with the focusing pattern preserved [Fig. \ref{fig:skyrmion_lens}(b)].

As the inverse process of partial convergence, rays emitted from the left focal point are partially converted into parallel rays after passing through the pseudo-magnetic field [Fig.~\ref{fig:skyrmion_lens}(c, d)]. 
Correspondingly, when spin waves emitted from a point source penetrate through the skyrmion, a substantial portion behaves like a plane wave, as demonstrated by spin wave flux profiles in micromagnetic simulations. 
The mutual transformation between point-source and plane-wave states in Figs.~\ref{fig:skyrmion_lens}(a-d) collectively showcase the magnonic lensing effect via an antiferromagnetic skyrmion.

\subsection{Magnonic lens set}

A single magnonic lens demonstrated in Fig. \ref{fig:skyrmion_lens}(a-d) can be straightforwardly extended to a coaxial lens set by assembling two skyrmions, as depicted in Fig.~\ref{fig:skyrmion_lens}(e, f).
By tuning the relative polarities of two skyrmions, the lens set switches between two functional modes: 
i) for identical skyrmion polarity, point-source spin wave emitting from the left focal point is partially refocused to the right focal point;
ii) for opposite skyrmion polarities, plane spin wave partially restores its original form after passing through a common focal point.

In the identical-polarity configuration [Fig.~\ref{fig:skyrmion_lens}(e)], spin waves excited from a point source at the left focal point of the first skyrmion sequentially encounter two skyrmions of identical polarity, the corresponding magnons experience duplicate pseudo-magnetic profiles and undergo deflection in the same direction, analogous to the magnonic lens set in Fig. \ref{fig:skyrmion_lens}(a) and (c). 
Consequently, divergent magnonic rays from the left focal point are converted to parallel rays, then converge again at the right focal point of the second skyrmion.
Such a focusing-collimation-refocusing scenario is validated by the spin wave flux profiles extracted from micromagnetic simulations.

In the opposite-polarity configuration [Fig.~\ref{fig:skyrmion_lens}(f)], a plane spin wave encounters two skyrmions of opposite polarity sharing a common focal point, the magnon rays are bent in opposing directions by inverse pseudo-magnetic profiles, resembling the combination in Figs. \ref{fig:skyrmion_lens}(a, d). 
This converts parallel magnonic rays to coaxial rays before reconversion to parallel propagation, and the corresponding collimation-focusing-recollimation scenario is again confirmed by micromagnetic simulations. 
Although a domain wall forms between domains hosting opposite-polarity skyrmions, its effect is limited to a slight lateral ray shift \cite{lan_geometric_2021} and is thus neglected.

\section{Discussions}

\subsection{Spatial patterns of  pseudo-magnetic fields}

\begin{figure}[tb]
	\centering
	\frame{\includegraphics[width=0.5 \textwidth]{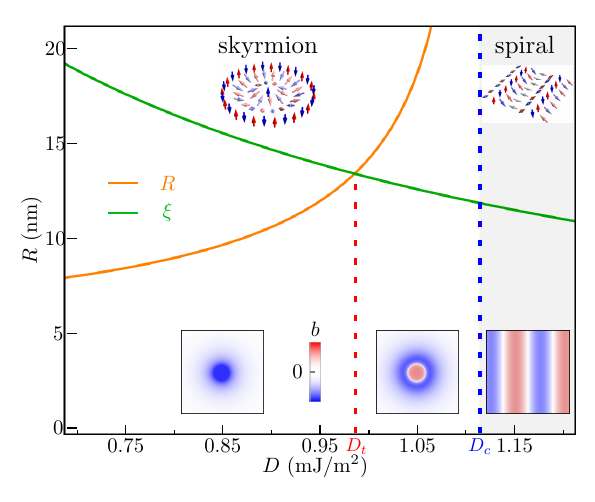}}
	\caption{
	{\bf Characteristic lengths and pseudo-magnetic field patterns as function of DM strength $D$.}
    Orange/green lines are for the skyrmion radius $R$ and the chiral length $\xi$.
	In upper insets, the red/blue arrows depict two sublattice magnetizations.
	In lower insets, the red/blue colors encodes the positive/negative values of the pseudo field.
	\label{fig:DM_threshold}
	}
\end{figure}

The pseudo-magnetic field generated by the magnetic skyrmion naturally segregates into three distinct patterns, demarcated by the threshold $D_t$ in \Eq{eqn:Dt} and the critical value $D_c$ in \Eq{eqn:R_D}, as illustrated in Fig.~\ref{fig:DM_threshold}. 
The emergence of an additional threshold $D_t$, alongside the critical value $D_c$, underscores the pivotal role of the chiral DM interaction in controlling pseudo-magnetic field and further the magnon deflections.

More explicitly, $D_c$ delineates the transition between radial and stripe  patterns of the pseudo-field, corresponding to the underlying magnetic texture evolving from  a magnetic skyrmion to a spin spiral \cite{rohart_Skyrmion_2013}.
Furthermore, the threshold  $D_t$ slightly lower than $D_c$,  distinguishes between uniform-sign and sign-reversal patterns, arising from the competitive interplay between the skyrmion radius $R$ and the chiral length $\xi$, as demonstrated in this work.

Owing to multiple approximations adopted in \Eq{eqn:b_0d_r0}, including the evaluations in radius $R$ and angle gradient $d\Theta/dr$, the exact value $D_t\approx\SI{0.94}{mJ/m^2}$ deviates slightly from the anticipated value  $D_t\approx\SI{0.98}{mJ/m^2}$ calculated from \Eq{eqn:Dt}, but the general partition scheme via $D_t$ and $D_c$ maintains.

\subsection{Focal length of a magnonic lens}
\label{sec:Focal}

\begin{figure}[tb]
	\centering
	\frame{\includegraphics[width=0.5 \textwidth]{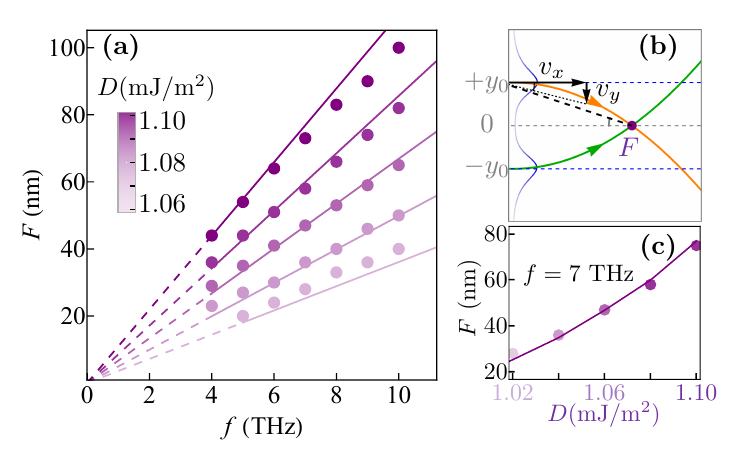}}
	\caption{
	{\bf Focal length of the skyrmion-based magnonic lens.}
	(a) Focal length as function of spin wave frequency $f$ under different DM strengths $D$.
	(b) Geometric relation between magnon trajectories and the focal point.
	Blue line represents the distribution of $B(y)$, orange/green arrows indicate magnon trajectories with pseudo-charges $\sigma=\pm 1$ passing through the upper and lower peaks, the intersection point corresponds to the focal point.
	(c) Focal length as a function of DM strength $D$  at a fixed frequency $f=\SI{7}{THz}$.
	In (a) and (c), the dots are extracted from micromagnetic simulations, and the solid line is based on \Eq{eqn:defl_F}. 
	\label{fig:focal_length}
	}
\end{figure}

The primary deflections of converging magnon rays occur at heights $y \approx \pm y_0$, corresponding to the locations where the path-integrated field $B(y)$ is maximal, as illustrated in Fig.~\ref{fig:skyrmion_lens}. Consequently, based on Fig.~\ref{fig:focal_length}(b), a geometric relation is established: $F / y_0 \approx \eta v_x / |v_y|$, where $F$ is the focal length. The correction factor $\eta$ accounts for curvilinear deviations in the magnon paths and is described by $\eta \approx \zeta (D - D_t)$, where $\zeta = \SI{0.34e4}{m^2/J}$ represents the modulation by the DM strength $D$, and $D_t$ is the threshold strength defined in \Eq{eqn:Dt}.

Based on above observations, the focal length $F$ is estimated as
\begin{align}
	\label{eqn:defl_F}
	F \approx \eta \frac{v_x y_0}{\frac{q_0}{m}B_0} 
	=\frac{\eta\hbar\omega}{q_0cB_0} y_0,
\end{align}
where $B_0 \equiv |B(y_0)|$ denotes the peak path-integrated field, and a linearized dispersion from \Eq{eqn:local_omega} is implicitly used. For a fixed skyrmion configuration, the focal length extracted from micromagnetic simulations exhibits a linear proportionality to the spin-wave frequency, as shown in Fig.~\ref{fig:focal_length}(a), demonstrating good agreement with the theoretical equation \eqref{eqn:defl_F}.

Increasing the DM strength $D$ affects the focal length through three simultaneous mechanisms:
i) an increase in the correction factor $\eta$;
ii) an enlargement of the skyrmion radius $R$, which elevates the characteristic height $y_0$;
iii) an enhancement of the chiral field $b^D$, which raises the peak field $B_0$.
Due to the synergistic scaling of $\eta$ and $y_0$, the magnon deflection intensifies at higher $D$, leading to a steeper slope in the $F$-$f$ relationship at elevated DM strengths, as evidenced in Fig.~\ref{fig:focal_length}(a).
 The dependence of the focal length $F$ on the DM strength $D$ is summarized in Fig.~\ref{fig:focal_length}(c) for a fixed frequency $f = \SI{7}{THz}$, where a monotonic increase is clearly identified.

\section{Conclusions}

In conclusion, we demonstrate that an antiferromagnetic skyrmion naturally functions as a magnonic lens when the Dzyaloshinskii-Moriya  strength exceeds a threshold. 
The underlying mechanism stems from magnon deflection by the DMI-induced pseudo-magnetic field, which exhibits opposite orientations in core versus shell regions. 
Utilizing magnetic skyrmions to manipulate spin-wave propagation direction,  establishes a new paradigm for designing purely magnetic information processing devices.

\acknowledgments
This work is supported by National Natural Science Foundation of China (Grant No. 11904260) and Natural Science Foundation of Tianjin (Grant No. 20JCQNJC02020).

\end{document}